\documentclass[11pt]{article}

\usepackage{amsfonts}
\usepackage{amsmath}
\usepackage{amssymb}
\usepackage{amsthm}

\usepackage{graphicx}

\usepackage{caption2}

\makeatletter \oddsidemargin-.25in \evensidemargin-.25in
\makeatother

\usepackage{graphicx}

\theoremstyle{plain}

\theoremstyle{definition}

\begin{document}
\title{Exceptional error minimization\\ in putative primordial genetic codes}

\author{Artem S. Novozhilov\footnote{e-mail: novozhil@ncbi.nlm.nih.gov} , and Eugene V. Koonin\footnote{e-mail: koonin@ncbi.nlm.nih.gov}\\[2mm]
{\small \textit{National Center for Biotechnology Information,}}\\ {\small \textit{National Library of Medicine, National Institutes of Health, Bethesda, MD 20894}}}

\date{}

\maketitle

\begin{abstract}
\noindent\textbf{Background:} The standard genetic code is redundant and has a highly non-random structure. Codons for the same amino acids typically differ only by the nucleotide in the third position, whereas similar amino acids are encoded, mostly, by codon series that differ by a single base substitution in the third or the first position. As a result, the code is highly albeit not optimally robust to errors of translation, a property that has been interpreted either as a product of selection directed at the minimization of errors or as a non-adaptive by-product of evolution of the code driven by other forces.\bigskip

\noindent{\textbf{Results:}} We investigated the error-minimization properties of putative primordial codes that consisted of 16 supercodons, with the third base being completely redundant, using a previously derived cost function and the error minimization percentage as the measure of a code's robustness to mistranslation. It is shown that, when the $16$-supercodon table is populated with 10 putative primordial amino acids, inferred from the results of abiotic synthesis experiments and other evidence independent of the code evolution, and with minimal assumptions used to assign the remaining supercodons, the resulting $2$-letter codes are nearly optimal in terms of the error minimization level.\bigskip

\noindent\textbf{Conclusions:} The results of the computational experiments with putative primordial genetic codes that contained only two meaningful letters in all codons and encoded 10 to 16 amino acids indicate that such codes are likely to have been nearly optimal with respect to the minimization of translation errors. This near-optimality could be the outcome of extensive early selection during the co-evolution of the code with the primordial, error-prone translation system, or a result of a unique, accidental event. Under this hypothesis, the subsequent expansion of the code resulted in a decrease of the error minimization level that became sustainable owing to the evolution of a high-fidelity translation system.\bigskip

\end{abstract}


\section{Background}
The standard genetic code, which is a mapping of 64 codons to 20 standard amino acids and the translation stop signal, is shared, with minor modifications only, by all life forms on earth (Woese, Hinegardner et al. 1964; Woese 1967; Ycas 1969; Osawa 1995). The apparent universality of the code implies that the last universal common ancestor (LUCA) of all extant life forms should have already possessed, together with a complex translation machinery, the same genetic code as contemporary organisms. One of the central principles of Darwinian evolution is that complex systems evolve from simple ancestors, typically if not always, via a succession of relatively small, incremental steps each of which increases fitness or at least does not lead to a decrease in fitness (Darwin 1859). In conformity with this continuity principle (Penny 2005; Wolf and Koonin 2007), it appears almost certain that the genetic code employed by the primordial translation system was substantially simpler than the modern code, which then evolved incrementally. The origin and evolution, if any, of the genetic code represent a major puzzle of modern biology; numerous hypotheses have been formulated but to date no generally accepted consensus has been reached (Knight, Freeland et al. 1999; Di Giulio 2005; Wong 2005; Novozhilov, Wolf et al. 2007; Higgs 2009; Koonin and Novozhilov 2009).

Several lines of evidence have been used to classify the standard 20 amino acids into `early' and `late' ones. The most straightforward indications, conceivably, come from experiments on abiogenic synthesis of organic molecules under supposedly realistic prehistoric atmosphere
conditions and external energy sources, a research direction pioneered by Miller and Urey in the
1950s (Miller 1953; Miller and Urey 1959; Miller, Urey et al. 1976).  The experiments of Miller and similar experiments subsequently performed by other groups under various models of the ancient atmosphere and using different energy sources, such as spark discharges, ultraviolet light, or irradiation with high energy charged particles (Kobayashi, Tsuchiya et al. 1990; Cleaves, Chalmers et al. 2008) yielded up to 10 standard amino acids (reviewed by Higgs and Pudritz, 2009). In general, the results of these experiments are remarkably coherent and lead to the same list of standard amino acids that can be produced under emulated primordial conditions:
\begin{equation}
	\textbf{\textrm{Gly, Ala, Asp, Glu, Val, Ser, Ile, Leu, Pro, Thr}}
\end{equation}

The second line of evidence is more speculative in nature and is based on the notion of the precursor-product pairs of amino acids. According to the coevolution theory of the genetic code, the present day amino acids that are used in translation are divided into two phases: phase I amino acids came from prebiotic synthesis, and phase II amino acids are entirely biogenic and were recruited into the code after the evolution of the respective biosynthetic pathways (Wong 1975; Wong 2005). Strikingly, the list of phase 1 amino acids that was derived from the analysis of biosynthetic pathways completely coincides with the above set of 10 amino acids observed in prebiotic amino acid formation experiments (Wong 1981). Furthermore, these ten amino acids have the lowest free energies of formation, an observation that is compatible with abiogenic emergence (Amend and Shock 1998; Higgs and Pudritz 2009).

Many attempts have been made to derive a universal order of the recruitment of amino acids during evolution (Trifonov 2000; Trifonov 2004). Using a combination of 60 different criteria, Trifonov reconstructed a `consensus temporal order of amino acids' (Trifonov 2004). Although this consensus order has been criticized on several grounds (Knight 2001), it should be noted that the resulting list of amino acids is in a nearly perfect agreement with the combined results of Miller and Urey type experiments. All amino acids synthesized under putative primordial conditions are classified as `early' in the consensus analysis, with one minor change: \textbf{Ile} is considered to be a `late' amino acid whose appearance is predated by \textbf{Arg} and \textbf{Asn} (see below).

On the strength of the consensus order, the results of Miller-type experiments, free energies of
formation, and the precursor-product relationship between amino acids, it seems most likely that,
although we generally cannot give an exact order of appearance of amino acids in the genetic
code, that the primordial genetic code should have coded for a subset of the present day amino
acid repertoire, and this subset, probably, included the 10 amino acids in list (1).

The genetic code is a mapping of the set of 64 codons onto the set of 20 standard amino acids
used in protein translation (and the stop signal). The continuity principle along with the classification of amino acids into early and late ones suggests that the primordial genetic code specified fewer amino acids than the universal standard code which immediately implies that the ancestral code was even more degenerate than the modern one. Importantly, there is essentially no doubt that, from the very emergence of the code, mRNAs (or, possibly, even chemically different primordial templates) were translated by triplets of nucleotides, even if only a few amino acids were encoded. Any speculation on a primordial code with singlet or doublet codons faces the apparently insurmountable obstacle of the subsequent code expansion to the present day triplet form, which obviously would be effectively fatal (Crick 1968). Furthermore, the three-base codon structure of the genetic code is likely to be determined by the physics of the interaction between monomers (Aldana, Cazarez-Bush et al. 1998; Aldana-Gonzalez, Cocho et al. 2003) and/or by possibility of simultaneous binding of two RNA adaptors on mRNA (Crick 1968; Travers 2006). If the code always consisted of triplets but specified 16 or fewer amino acids, it appears likely that only the first two bases of each codon were informative in the primordial code whereas the third base did not contribute to coding. In other word, the primordial mRNA sequences would have the form \verb"XYNXYNXYN…" where \verb"X, Y" are `meaningful' nucleotides, and \verb"N" stands for any nucleotide (Woese 1965; Patel 2005; Wu, Bagby et al. 2005; Travers 2006). That the primordial code would have this particular organization is strongly suggested by the structure of the extant code in which redundancy is concentrated almost entirely in the third base; apparently, it is the first and, especially, the second bases that ensure the stability of the interaction between codons and cognate anticodons (Travers 2006).

It is therefore not unrealistic to propose that the primordial genetic code consisted of 16 supercodons (4-codon series, \verb"XYN") and encoded 16 or fewer amino acids, possibly, the 10 inferred early amino acids listed above (1). Here we investigate the properties of such putative primordial codes and show that, under some additional, simple assumptions, they would possess extraordinary error minimization properties.

\section{Results and Discussion}
Assuming that, at a particular early stage of evolution, the primordial genetic code consisted of 16 supercodons, we postulate the following `parsimony principle':\bigskip

\textit{If the primordial code encoded an amino acid, then this amino acid was encoded by the same supercodon (four-codon series) that encodes the same amino acid in the standard genetic code (or, at least, a subset of the series encodes the same amino acid).}\bigskip

The expansion of the code from codons with two meaningful letters to codons with three meaningful letters is required to involve the minimum possible number of amino acid reassignments; accordingly, expansion of the code only allows recruitment of a subset of codons in a supercodon for a new amino acid but not reassignment of codons within the primordial set of amino acids. This assumption is natural because reassignment of amino acids between supercodons series, obviously, is substantially more disruptive than capturing new amino acids within pre-existing codon series (Higgs 2009). With one exception, there are no contradictions between the list of putative ancestral amino acids (1) and the parsimony principle: most of the `early' amino acids are encoded by four-codon series, and only two, \textbf{Asp} and \textbf{Glu}, do not satisfy the two-letter code scheme and the parsimony principle in that they are encoded by the same supercodon. Following the suggestion of Travers (Travers 2006), we speculate that decoding of the supercodon \verb"GAN" initially was stochastic, that is, these very similar amino acids were incorporated more or less randomly in response to the codons of this series, and differentiation of \textbf{Asp} and \textbf{Glu} was established only after the expansion of the genetic code to three-letter codons.

Using the parsimony principle, the primordial two-letter code can be partially reconstructed as shown in Fig. 1. Obviously, the parsimony principle does not allow one to infer the assignment for those supercodons that, in the standard code, do not encode any of the primordial amino acids (question marks in Fig. 1). To fill these gaps, additional assumptions on the amino acid assignments are required.
\begin{figure}
\centering
\includegraphics[width=0.6\textwidth]{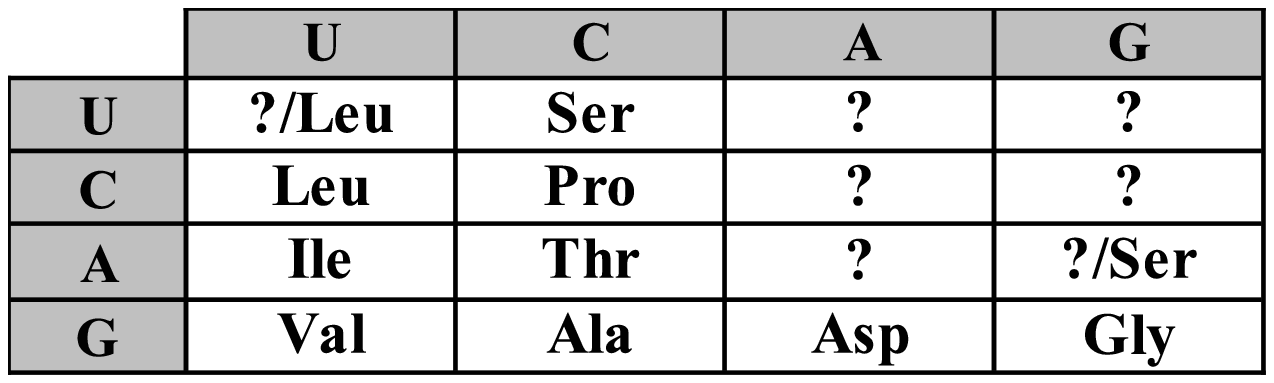}
\caption{A 2-letter code consisting of 16 supercodons with the assignment inferred from the list of `early' amino acids (1) and the parsimony principle. }
\end{figure}

It is instructive to compare the putative core of the primordial genetic code in Fig. 1 with the order of stabilities of the interactions between the first two bases of codons and the cognate anticodons (Travers 2006) (Fig. 2). There is a striking congruence between the two lists of amino acids. Indeed, the supercodons for 10 early amino acids include 9 of the top 10 most strongly interacting dinucleotides as determined by the stacking and melting thermostabilities. The sole exception is the supercodon \verb"CGN" that encodes \textbf{Arg}, not an early amino acid, but is more stable than \verb"CUN" and \verb"AUN", which encode the early amino acids \textbf{Leu} and \textbf{Ile}, respectively (Fig. 2).

\begin{figure}
\centering
\includegraphics[width=0.6\textwidth]{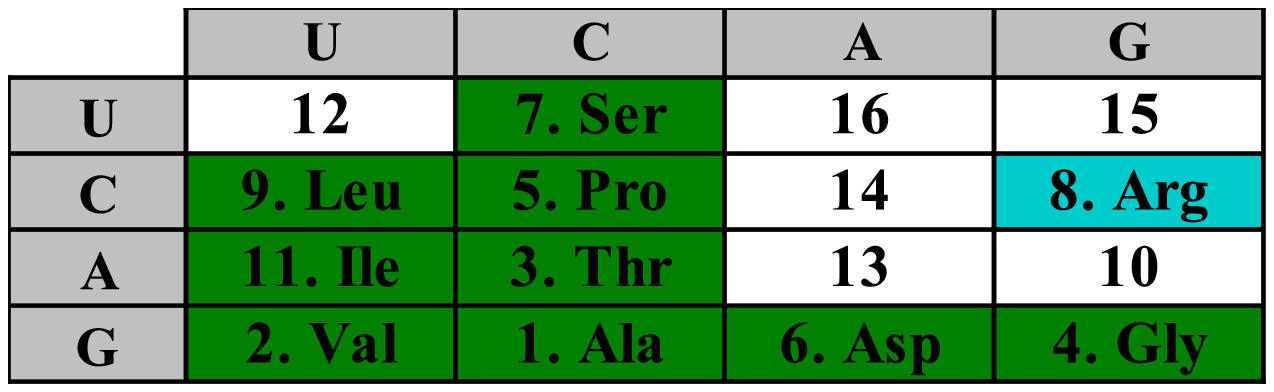}
\caption{The order of stabilities of base interaction in the first two codon positions of the standard code according to Travers (Travers 2006). The green highlighting shows the cells that correspond to the supercodons encoding the `early' amino acids (1). The only difference from the list (1) is shown in blue.}
\end{figure}
The standard genetic code is manifestly non-random. In particular, the assignments of amino acids to codons are such that the detrimental effect of mistranslation and/or mutation is minimized. That is, in the standard genetic code, codons that differ by one nucleotide code for physicochemically similar amino acids, thus reducing the cost of possible mistranslations and mutations. Quantitative evidence in support of this error-minimization property comes from the comparison of the standard code with random alternatives (Haig and Hurst 1991; Freeland and Hurst 1998; Gilis, Massar et al. 2001; Novozhilov, Wolf et al. 2007). It is thus necessary, when considering any scenario for the origin and evolution of the code, to account for this property. There are two possible explanations for error minimization in the code. The first possibility is that the high degree of error minimization is a byproduct of other processes that shaped the structure of the genetic code, e.g., (Stoltzfus and Yampolsky 2007; Massey 2008; Higgs 2009). The alternative is the error-minimization (adaptive) theory of the code evolution which posits that the code evolved under the selective pressure to reduce the consequences of mistranslations and/or mutations (Freeland, Knight et al. 2000). Here we use the same quantitative approach ((Novozhilov, Wolf et al. 2007) and see Methods for details) to estimate the error-minimization level of the putative primordial `two-letter' codes that have at their core the amino acid assignments shown in Fig. 1.

For the time being, let us disregard the unassigned entries in the code table (question marks in Fig. 1). For any permutation of the amino acid assignments in the code table, a code cost can be calculated. This cost depends on the probability of a given mistranslation error and on the relative cost associated with the replacement of the corresponding wild-type amino acid with a new one (see Methods for the exact details of the calculation of the code cost). Disregarding the unassigned supercodons but otherwise allowing all permutations of amino acid assignments within the rest of the supercodons (9, 10 or 11, depending on whether amino acids are assigned to the \verb"UUN" and \verb"AGN" supercodons or not), we find that the code structure in Fig. 1 is close to optimal in terms of error minimization. More precisely, the code structure in Fig. 1 is extremely robust to translational errors irrespective of the assignments of the \verb"UUN" or \verb"AGN" supercodons. In two of the four possible cases (Fig. 3a and 3d), there is no permutation that would reduce the cost of the code, that is, the minimization percentage ($MP$; see Methods for details) of the code is 1; in the other two cases, the optimal codes differ from the code in Fig. 1 only by permutations in the second column, and $MP$ of these codes is greater than 0.98 (Figs. 3b and 3c).
\begin{figure}[tbh]
\centering
\includegraphics[width=0.95\textwidth]{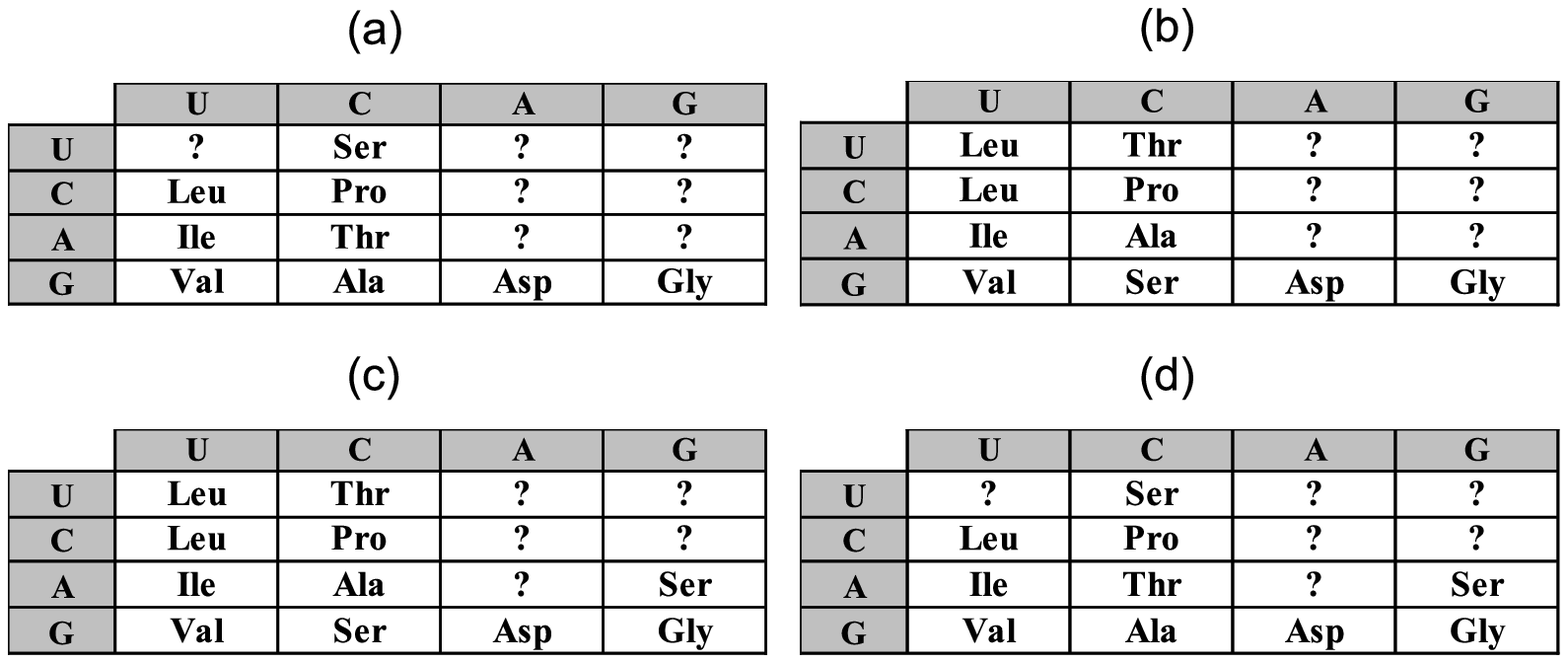}
\caption{Error minimization level of 2-letter codes. Starting from a random permutation of amino acids assignments within the supercodons without question marks, optimization was performed to find the least costly assignment according (see Methods). $(a)$ 9 amino acids, the optimum coincides with the code in Fig. 1, $MP=1$; $(b)$ additional assignment of \textbf{Leu} is included; the optimum differs from the code in Fig. 1 by permutations in the second column, $MP=0.986$; $(c)$ \textbf{Leu} and \textbf{Ser} are added, $MP=0.985$; $(d)$ only \textbf{Ser} is added, the optimum is the same code as in Fig. 1, $MP=1$.}
\end{figure}

One possible interpretation of the high robustness of the doublet codes shown in Fig. 3 could be that, with this particular choice of amino acids and supercodons, and the employed measure of the code cost, most of the random codes yield low cost. However, this is not the case, as can be seen from the distribution of random code costs shown in Fig. 4, for the versions of the code from Figs. 3a and 3d. Interestingly, the cost distribution for the code from Fig. 3a is bimodal (a similar distribution was obtained for the code in Fig. 3b; not shown) whereas the distribution for the code from Fig. 3d is a more typical, roughly bell-shaped one. The difference between the cost of the standard code (Fig. 1) and the means of the distributions measured in standard deviations is 2.2, 2.65, 2.91, and 2.5 for the cases $(a)$, $(b)$, $(c)$, $(d)$ in Fig. 3, respectively. Even in the cases $(b)$ and $(c)$, where the assignment of amino acids to supercodons could be improved, the code structure in Fig. 1 is extremely close to the optimum (that, the global cost minimum).
\begin{figure}[bh!]
\centering
\includegraphics[width=0.95\textwidth]{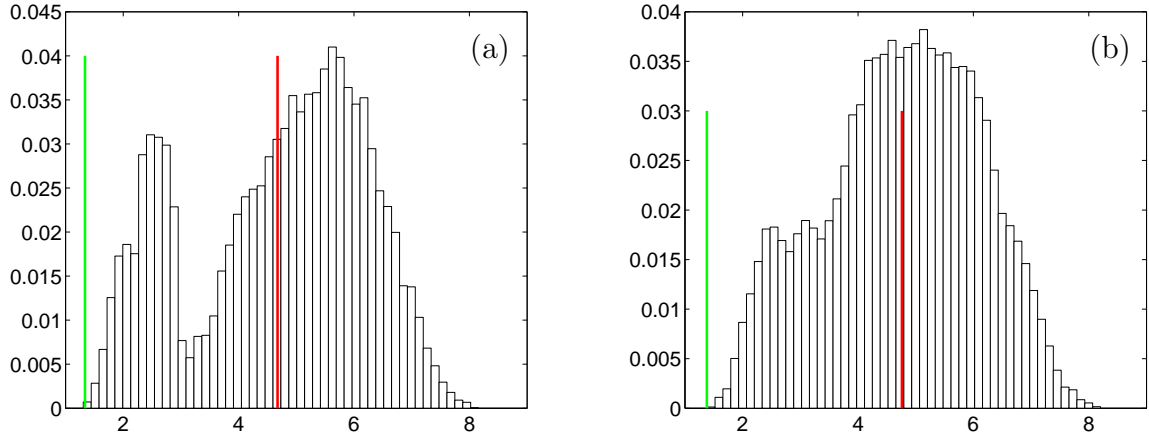}
\caption{Distributions of random code costs obtained by permutation of amino acid assignments to the supercodons without question marks in Fig. 2. The green line shows the cost of the code from Fig. 1, and the red line shows the mean of the distribution. $(a)$ 9 amino acids from Fig. 2a are used, the distance from the mean to the red line is 2.2 standard deviations; $(b)$ \textbf{Ser} is added to the list of the 9 amino acids, the distance from the mean to the red line is 2.5 standard deviations.}
\end{figure}

Thus, we showed that the part of the putative two letter primordial genetic code that can be unambiguously inferred assuming the list of early amino acids (1) and the parsimony principle is, in effect, optimal with respect to error minimization property. It seem virtually impossible to explain away this `perfect' structure as a by-product of some evolutionary process for which error minimization is of secondary importance or neutral. Neither is it possible to explain these codon assignments by random effects because, for instance, for the code in Fig. 3a, there are $181440=9!/2$ alternatives all of which are worse than the one shown in the figure.

\begin{figure}
\centering
\includegraphics[width=0.6\textwidth]{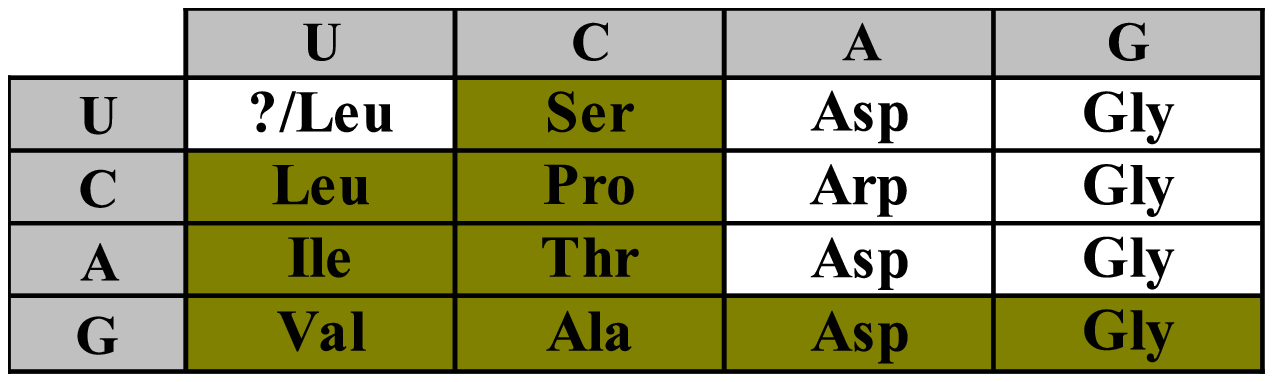}
\caption{A 2-letter code consisting of 16 supercodons assigned according to the list of the `early' amino acids (1), the parsimony principle, and the `4-column' theory. Dark green cells are those that are assigned in Fig. 1, and question marks in Fig. 1 are replaced with amino acid assignments from the respective columns}
\end{figure}

There is, of course, a major caveat in these conclusions. The code cost function is not linear in the sense that adding another amino acid generally destroys the optimal assignments. Given that we disregarded some of the supercodons when performing the numerical experiments described above, the observed extreme error minimization of the putative primordial 2-letter code might be illusory. Therefore, additional assumptions were necessary to fill those supercodons of the 2-letter codes which do not have amino acid assignments after applying of the parsimony principle to the standard code given the list of early amino acids (1). A possible solution that we consider first, is to fill unassigned cells with the amino acids from the same column, in accordance with the 'four-column' theory of the origin of the genetic code (Woese 1965; Higgs 2009). For instance, consider the code in Fig. 5. We take the amino acid assignments from Fig. 1 whenever possible, disregard \textbf{Ser} for supercodon \verb"AGN", so that the whole column codes for the same amino acid, and either assign \textbf{Leu} to \verb"UUN", because the closest amino acid in this code is \textbf{Leu}, or assume the existence of two supercodons for \textbf{Leu} (incidentally, the most abundant amino acid in extant proteins) already at the 2-letter stage of the code evolution. Allowing random permutations of amino acid assignments within the colored cells in Fig. 5 and filling other cells using the `column-wise' approach, the error minimization properties of the code in Fig. 5 can be assessed. It turns out that the code in Fig. 5 is also highly robust although not quite at the level of the abridged codes in Fig. 3 (Fig. 6). Specifically, if supercodon \verb"UUN" is filled using the assignment of \verb"CUN" (\textbf{Val}), the $MP$ of the code from Fig. 5 is 0.94 (Fig. 6a); if two supercodons for \textbf{Leu} are assumed, then the $MP$ is 0.987, and the optimal code is very close to that in Fig. 5 (Fig. 6b). In both cases, lowest cost was obtained for the assignments where the third and fourth columns code for \textbf{Asp} and \textbf{Gly}, respectively. The distributions of the random code costs are show in Fig.~7.
\begin{figure}[th]
\centering
\includegraphics[width=0.95\textwidth]{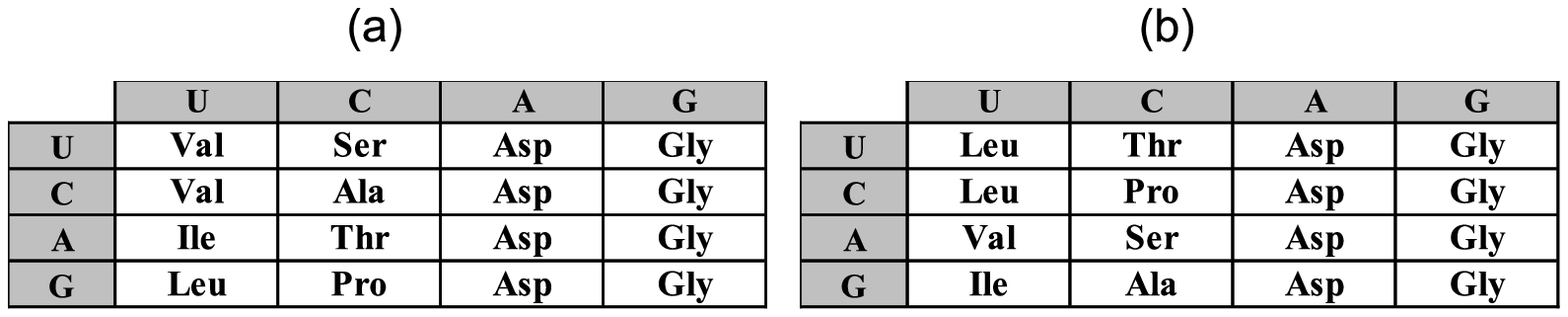}
\caption{Error minimization level of 2-letter codes. Starting from a random permutation of amino acids assignments among the supercodons highlighted in Fig. 5, optimization was performed to find the least costly assignment. $(a)$ 9 amino acids, the optimum code is shown, $MP=0.94$; (b) The number of supercodons coding for \textbf{Leu} is fixed to 2; $MP=0.987$}
\end{figure}

\begin{figure}[bh!]
\centering
\includegraphics[width=0.95\textwidth]{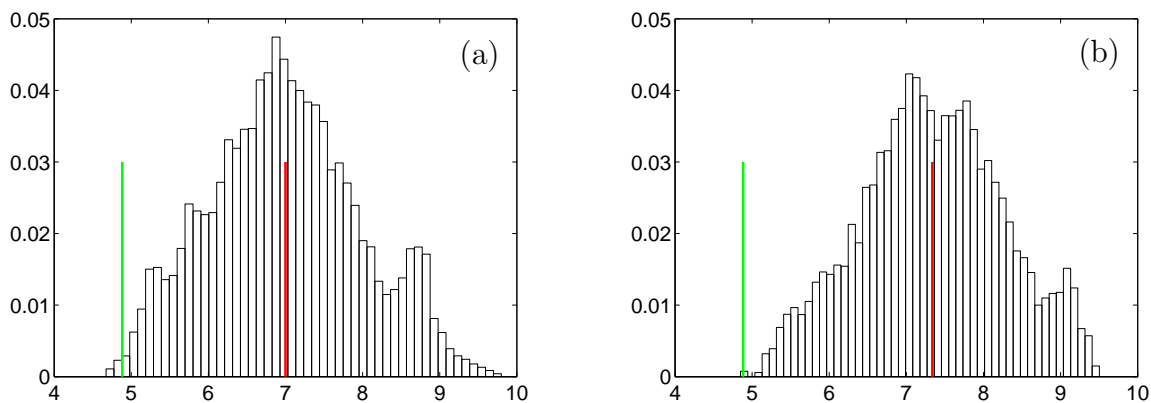}
\caption{The distributions of random code costs for the experiments in Fig. 6. The green line shows the cost of the code from Fig. 5, the red line shows the mean of the distributions. $(a)$ 9 amino acids from Fig. 5, the distance from the mean to the red line is 2.16 standard deviations; $(b)$ Two supercodons assigned for \textbf{Leu}; the distance from the green line to the mean is 2.6 standard deviations.}
\end{figure}

Thus, at least, the part of the 2-letter code that can be inferred from the standard code using the set of (putative) primordial amino acids, the parsimony principle, and a straightforward additional assumption for the assigning the remaining supercodons, is structured in such a way that an a priori chosen standard cost function (see Methods) renders the code near-optimal. Indeed, the most conservative estimates yield $MP > 0.98$ for the cases when the question marks Fig. 1 are disregarded, and $MP>0.94$ when the `four-column' theory is used to assign amino to the unassigned supercodons (Fig. 6), in a sharp contrast to the 78\% $MP$ for the standard code (Koonin and Novozhilov 2009) (this estimate was obtained using the same cost function as described in the Methods section but for the complete, standard genetic code, and is somewhat higher than the previously reported estimates (Di Giulio, Capobianco et al. 1994)).

A different approach to assigning the vacant supercodons in the 2-letter in Fig. 1 involves using the parsimony principle not only for the putative early amino acids but for all supercodons. Under this strategy the 2-letter codes cease being special with respect to error-minimization. Consider, for instance, the code shown in Fig. 8a that obtained from the standard code using the parsimony principle. This version of the 2-letter code was proposed as a possible ancestral code (Copley, Smith et al. 2005) and was analyzed with respect to error minimization (Butler and Goldenfeld 2009). This code has $MP$ of 0.51, and the result does not change qualitatively when ambiguous amino acid assignments are changed (for instance, when \textbf{Gln} is substituted for \textbf{His}). Here our conclusion is in agreement with the conclusions of (Butler and Goldenfeld 2009) that were obtained using a different cost function.

With regard to the low error minimization in 2-letter codes obtained using the parsimony principle, we were interested in determining which amino acid assignments contributed the most to this non-optimality. In the standard genetic code, the most non-optimally assigned amino acid is \textbf{Arg} (Novozhilov, Wolf et al. 2007); the underlying reason is not only the placement of \textbf{Arg} in the code table as such but also the fact that \textbf{Arg} has 6 codons and so makes a disproportionate contribution to the cost of the code. In 2-letter codes, an amino acid can be encoded by two supercodons at the most, so it would not be surprising if an amino acid(s) other than \textbf{Arg }occupied the `worst' position from the point of view of the error minimization

\begin{figure}
\centering
\includegraphics[width=0.95\textwidth]{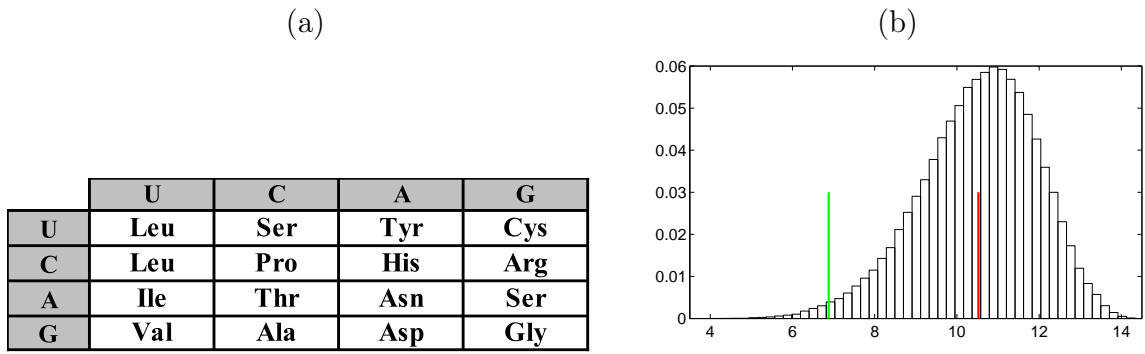}
\caption{Error minimization level of 2-letter codes. $(a)$ A 2-letter code obtained using the parsimony principle. For the cells with an ambiguous assignment, one random amino acid is chosen; $(b)$ The distribution of the costs of the random 2-letter codes obtained by permutation of amino acid assignments in $(a)$, the green lines shows the cost of the code from $(a)$ and the red line shows the mean; $MP= 0.51$, the distance from the mean is 2.6 standard deviations.}
\end{figure}

To address this question for 2-letter codes but taking into account all 20 standard amino acids, we devised the following experiment: for a given natural number $N$, choose randomly $N\leq16$ cells in the 16-cell code table. Then assign amino acids to the chosen cells according to the parsimony principle (if for some cells two amino acids are encoded in the respective 4-codon series in the standard code, one is randomly chosen). Allowing permutations of amino acid assignments between these fixed   cells, we can estimate the $MP$ for the given code. Other cells, not chosen in the experiment, can be disregarded, as it was done for the code in Fig. 1, or filled by using, e.g., the four-column rule specified above, as in Fig. 5. Repeating this procedure and collecting random codes with high $MP$, we can rank the amino acids by the frequency with which they are found in highly optimized codes and similarly rank the cells (supercodons) in the code table.

Independent of the number of chosen cells $N$ and the strategy that is used to fill (or not to fill) the remaining cells, the results qualitatively appear as shown in Fig. 9. The general conclusion is that the major reason of non-optimality of 2-letter codes obtained with the parsimony principle (as in Fig. 8a) are the amino acid assignments in the supercodons \verb"UAN" and \verb"UAG" which correspond to \textbf{Tyr}, \textbf{Cys}, and \textbf{Trp} (and two of the three the stop codons) in the standard code. We were unable to discriminate the effects of other amino acids except that these effects were relatively small and sensitive to the choice of $N$ (Fig. 9 and data not shown) but the non-optimality of the assignments of \textbf{Tyr}, \textbf{Cys}, and \textbf{Trp} was striking and is unambiguous (Fig. 9).
\begin{figure}[t!]
\centering
\includegraphics[width=0.85\textwidth]{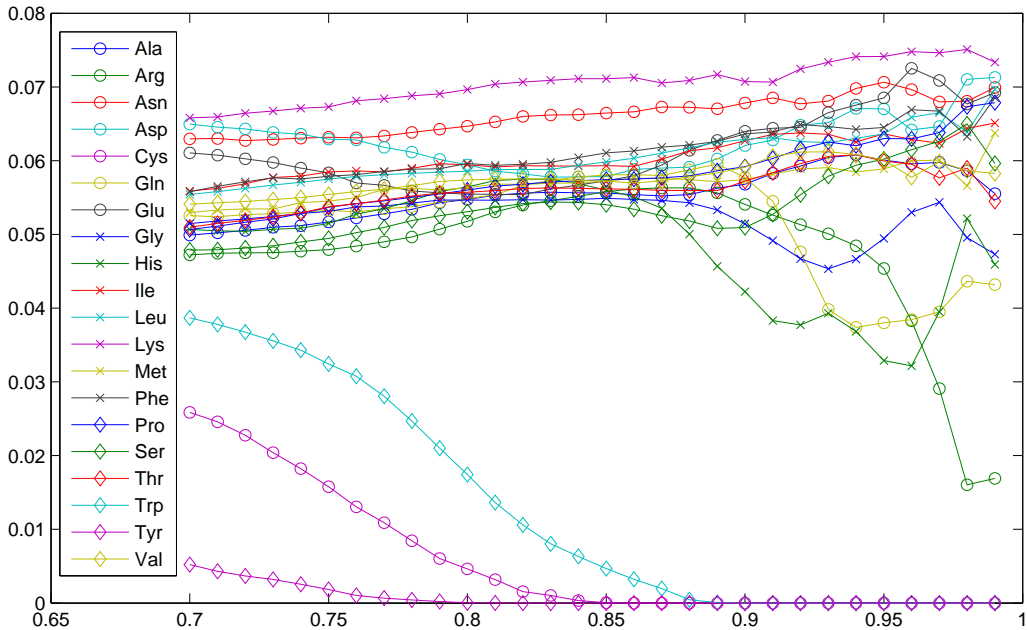}
\caption{The effects of individual amino acid assignments on the error minimization level of 2-letter codes similar to the code in Fig. 8a.  Amino acid frequencies ($y$-axis) are shown for the random codes (see text for details) for which $MP$ is equal to or higher than the given value ($x$-axis). For instance, there are no codes containing \textbf{Trp} which with $MP > 0.9$, and no codes containing \textbf{Tyr} with $MP > 0.8$.}
\end{figure}

Taking into account that \textbf{Tyr}, \textbf{Cys}, and \textbf{Trp} are among the `latest' amino acids according to Trifonov's consensus of amino acid appearance (Trifonov 2004), and that they are coded by supercodons with the lowest stability of codon-anticodon interactions (Fig. 2), it appears most likely that the primordial 2-letter genetic code did not accommodate these amino acids that  were added to the amino acid repertoire only after the transition to the standard 3-letter code. Given these observations, we assessed the error minimization level of 2-letter codes without assigning the supercodons \verb"UAN" and \verb"UGN" (Fig. 10). Such a 2-letter code is significantly more robust than the fully specified code in Fig. 8a: the $MP$ of this code is 0.88, a value that is significantly greater than the $MP$ of the standard code (0.78), with the probability to find a better code of approximately $1/50000$.

\begin{figure}[th!]
\centering
\includegraphics[width=0.95\textwidth]{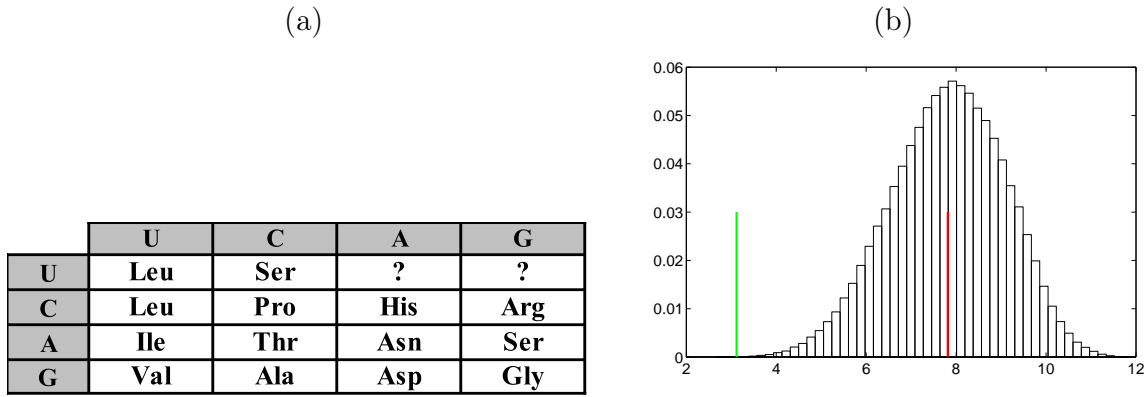}
\caption{Error minimization levels of 2-letter codes. $(a)$ A 2-letter genetic code obtained using the parsimony principle. For the cells with ambiguous assignment, one random amino acid is chosen; two supercodons, UAN and UGN, are disregarded; $(b)$ The distribution of the costs of random 2-letter codes obtained by permutation of amino acid assignments in $(a)$, the green line shows the cost of the code in $(a)$, and the red line shows the mean; $MP=0.88$, the distance from the mean is 3.7 standard deviations.}
\end{figure}

In the original experiment on spontaneous formation of organic compounds, Miller (1953) observed detectable amounts of only three amino acids: \textbf{Ala}, \textbf{Asp} and \textbf{Gly}. In most of the subsequent abiogenic synthesis experiments, these amino acids were most abundant. Thus, it seems to be a plausible assumption that these amino acids were the first to be encoded unambiguously in the primordial code, and their positions were fixed by chance (`frozen accident' \textit{sensu} Crick). We measured the level of error minimization for the 2-letter code, with permutations of amino acid assignments allowed only for the entries other than \verb"GCN", \verb"GAN", \verb"GGN", \verb"UAN", and \verb"UGN" (Fig. 11a).
\begin{figure}[bh!]
\centering
\includegraphics[width=0.95\textwidth]{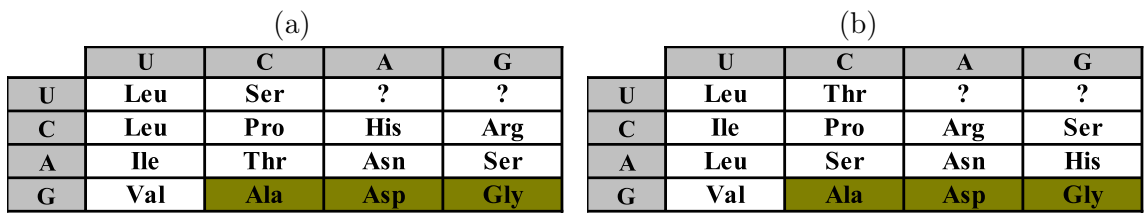}
\caption{Error minimization levels of 2-letter codes. $(a)$ A 2-letter code similar to that in Figure but with fixed assignments for 3 amino acids (dark green); $(b)$ optimal code found by permutation of amino acid assignments in $(a)$, $MP=0.91$. }
\end{figure}

The codes in this group are not exceptionally robust to translational mistakes ($MP$ is 0.91-0.93 depending on the choice of amino acids for the \verb"UUN", \verb"CAN", \verb"AAN", \verb"AGN" supercodons). Inspection of the optimal codes readily reveals the main source of this non-optimality: in all optimal solutions \textbf{Arg} changes its position from the fourth to the third column of the table (Fig. 11b). Arginine has a prominent place in the study of the genetic code evolution. From the point of view of the adaptive theory, \textbf{Arg} is the amino acid that brings most non-optimality into the standard code (Jukes 1973; Tolstrup, Toftgard et al. 1994; Novozhilov, Wolf et al. 2007). At the same time, \textbf{Arg} is the amino acid for which the strongest support for a stereochemical affinity with the respective codon is available (Knight and Landweber 1998; Knight and Landweber 2000; Knight, Landweber et al. 2003; Yarus, Caporaso et al. 2005).

Having found that the position of \textbf{Arg} is so critical for the code robustness, the following experiment was conducted. We start with the code table in Fig. 10a and the contribution of the \verb"UAN" and \verb"UGN" supercodons disregarded. From all other cells, two amino acids are chosen randomly and their assignments are fixed. Thus, a code table is obtained in which 4 cells are fixed (the two chosen amino acids and the supercodons \verb"UAN" and \verb"UGN"), whereas the assignments for the remaining 12 cells are freely permuted, and the $MP$ is calculated for all such permutations. We found that \textbf{Arg} is unique in this setting: for most of the amino acids, pairing with \textbf{Arg} yields the highest $MP$ of all possible pairings. The resulting $MP$ values are all within the range of 0.89 to 0.94, with one notable exception: if the pair \textbf{Asp}-\textbf{Arg} is fixed, then $MP$ of the code in Fig. 12a is 0.98 (the optimal code is shown in Fig. 12b).
\begin{figure}[t!]
\centering
\includegraphics[width=0.95\textwidth]{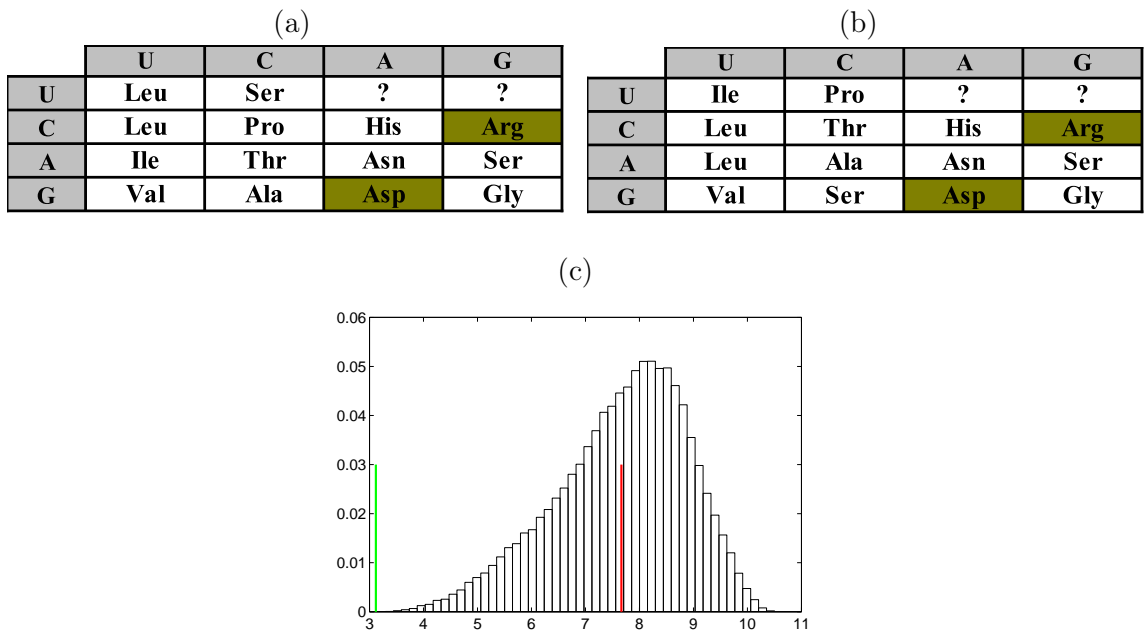}
\caption{Error minimization levels of 2-letter codes. $(a)$ A 2-letter code obtained using the parsimony principle, with two fixed amino acid assignments shown by dark green highlighting; $(b)$ the result of optimization of the code by permutation of amino acid assignments in $(a)$; $(c)$ the distribution of the costs of random codes, the green line shows the cost of the code in $(a)$, and the red line shows the mean; $MP=0.98$, the distance from the mean is 3.75 standard deviations}
\end{figure}

\section{Conclusions}
Immediately after the standard genetic code was deciphered, it has become apparent that the code table has a distinctly non-random structure, with similar codons encoding related amino acids (Woese 1965; Rumer 1966; Vol'kenshtein and Rumer 1967; Woese 1967). An obvious and crucial question is, what are the underlying causes of this regularity?

Three major conceptual frameworks have been developed to explain the regularities in the code (Knight, Freeland et al. 1999; Knight, Freeland et al. 2001; Koonin and Novozhilov 2009). The error-minimization theory holds that the structure of the code is the result of selection for robustness to mistranslation (Freeland and Hurst 1998; Freeland, Knight et al. 2000; Freeland, Knight et al. 2000; Freeland, Wu et al. 2003; Novozhilov, Wolf et al. 2007). The stereochemical theory posits that the code is determined, mostly, by stereochemical affinities between coding triplets (codons and/or anticodons) and the cognate amino acids (Yarus 1991; Yarus 2000; Yarus, Caporaso et al. 2005). The stereochemical theory alone cannot account for the high level of error minimization in the standard code; moreover, the affinity between cognate triplets and amino acids appears to be largely independent of the highly optimized amino acid assignments code (Caporaso, Yarus et al. 2005). To explain the structure of the code, the proponents of the stereochemical theory postulate that only part of the code is stereochemically fixed, whereas other amino acid assignments are free to be redistributed, and reassignment of even a few amino acids is sufficient for substantial optimization of the code (Yarus, Caporaso et al. 2005). The third, coevolution theory postulates that the structure of the code reflects the biosynthetic pathways of amino acid formation. Under this scenario, during the code evolution, subsets of codons for precursor amino acids have been reassigned to encode product amino acids (Wong 1975; Wong 2005). Then, the high level of the code optimality is just a byproduct of the evolutionary expansion in the code, and selection for robustness played only a minor role in the evolutionary shaping of the code (Stoltzfus and Yampolsky 2007) (although this role is still maintained to be more important than that of stereochemical affinities (Wong 2007)). Recently, a detailed extension of the coevolution theory has been developed where the evolutionary steps of the genetic code evolution are given in details (Di Giulio 2008).

A common feature of the stereochemical theory and the coevolution theory that is central to the present study is that the level of error minimization in the primordial codes is assumed to be low but is thought to have increased to the present level as a results of late amino acid reassignments (stereochemical theory) or capture of new amino acids (coevolution theory). The results of the present analysis of 2-letter codes are at odds with this view. Under minimal additional assumptions about the primordial code, which include the lack of unique assignments for the supercodons \verb"UAN" and \verb"UGN", and the assignment of \textbf{Arg} on the basis of stereochemistry, we arrive to the conclusion that the primordial 2-letter code was either shaped almost exclusively by the selective forces to minimize the impact of mistranslation or emerged in this highly robust form as a result of an extremely rare event. Indeed, with these assumptions only, combined with a random fixation of the assignment for \textbf{Asp}, we find that the 2-letter code constructed from the putative primordial amino acids using the parsimony principle is nearly optimal with respect to error minimization ($MP>0.98$).

We suspect that the high robustness of the primordial code is a pre-requisite for the evolution of the translation system that was, probably, considerably more error-prone at the early stages of evolution than it is in modern organisms (Noller 2004; Noller 2006; Wolf and Koonin 2007). The subsequent expansion of the code, whether it occurred on a stereochemical basis or by coevolution led only to a decrease of the code robustness. This course of evolution was made possible by the evolution of the modern, high-fidelity translation system as well as proteins that are partially optimized for robustness to misfolding (Drummond and Wilke 2008), and was driven by the selective advantage of the increased diversity of the amino acid repertoire.

\section{Methods}
\subsection*{The code cost}
The genetic code is a mapping $a\colon C\to A$  that assigns an amino acid (or stop signal) $a(c)\in A$ for any codon $c$. The cost function can be written as
$$
\varphi(a(c))=\sum_{c}\sum_{c'}p(c'|c)d(a(c),a(c')),
$$
where matrix $p(c'|c)$ gives the probability of misreading codon $c$ as codon $c'$. The numerical values for this matrix can be obtained in different ways. We adopt the most commonly used scheme where only codon pairs that differ in one nucleotide are considered. To account for the transition-transversion bias at the levels of both mutation and translation, transitions are set to be two-fold more frequent than transversions in the first position of the codon, and five-fold more frequent in the second position. Specifically, we use the matrix
$$
p(c|c')=\begin{cases}
1/N&\mbox{ if $c$ and $c'$ differ in the 1st base only and cause a transision}\\
0.5/N&\mbox{ if $c$ and $c'$ differ in the 1st base only and cause a transversion}\\
0.5/N&\mbox{ if $c$ and $c'$ differ in the 2nd base only and cause a transision}\\
0.1/N&\mbox{ if $c$ and $c'$ differ in the 2nd base only and cause a transversion}\\
0&\mbox{ otherwise}
\end{cases}
$$
where $N$ is a constant to ensure that $\sum_c p(c'|c)=1$. This assumption was widely used in previous analyses of the error minimization in genetic codes (see (Freeland and Hurst 1998; Gilis, Massar et al. 2001; Novozhilov, Wolf et al. 2007) for discussion). It could be argued that recent experimental evidence shows no bias in the second base (Kramer and Farabaugh 2007; Higgs 2009)  but more data is needed for reliable conclusions, so we adhered to the traditional scheme that was chosen a priori, before performing  any simulations and numerical experiments.

The matrix $d(a(c),a(c'))$ defines the cost of replacing amino acid $a(c)$ with amino acid $a(c')$. The choice here is also manifold (see, e.g., (Higgs 2009), where a complex index is defined to estimate the cost of amino acid replacement) but we employed only one measure of amino acid similarity, namely the Polar Requirement Scale proposed by Woese et al. (Woese, Dugre et al. 1966), which is a measure of amino acid hydrophobicity. The cost of replacement of one amino acid with another is calculated as $d(a(c),a(c'))=(p(a(c))-p(a(c')))^2$ , where $p(a(c))$  and $p(a(c'))$ are the values of the amino acids $a(c)$ and $a(c')$ at the Polar Requirement Scale.
Using this formalism, the cost of any genetic code can be calculated; the smaller the value, the higher the error minimization level of the given code.

\subsection*{Minimization percentage of a code}
To estimate the relative level of robustness to translational errors for a given code, we calculate Minimization Percentage ($MP$):
$$
MP=\frac{E(\varphi)-\varphi_{code}}{E(\varphi)-\varphi_{opt}}\,,
$$
where  $E(\varphi)$ is the mean value of the distribution of code costs which are obtained as permutations of the amino acid assignments in the code table, $\varphi_{code}$ is the cost of the given code, and $\varphi_{opt}$ is the cost of the optimal code which can be obtained for the given set of amino acids; the criterion of optimality is robustness to translational mistakes. To find the optimal code and its cost, exhaustive search of some of the two-letter codes is possible but this search is computationally intensive for the numerous 2-letter codes that we analyzed in the course of this
study. In most cases, a heuristic combinatorial algorithm was used (the Great Deluge Algorithm by Dueck, 1993), 3 to 5 solutions were identified, and the solution with the lowest cost was taken as the optimal.

\paragraph{Author contributions:}ASN and EVK conceived of the study, ASN performed the computational experiments and wrote
the original draft, EVK added the biological interpretation and wrote the final version of the
article that was approved by both authors.

\paragraph{Acknowledgements:}The authors' research is supported by intramural funds of the DHHS (NIH, National Library of
Medicine.\\[2mm]

\textbf{\Large{References}}\\[1mm]
\small

Aldana-Gonzalez, M., G. Cocho, et al. (2003). "Translocation Properties of Primitive Molecular Machines and Their Relevance to the Structure of the Genetic Code." Journal of Theoretical Biology 220(1): 27-45.\\[-3mm]

Aldana, M., F. Cazarez-Bush, et al. (1998). "Primordial synthesis machines and the origin of the genetic code." Physica A 257(1): 119-127.\\[-3mm]

Amend, J. P. and E. L. Shock (1998). "Energetics of amino acid synthesis in hydrothermal ecosystems." Science 281(5383): 1659--1662.\\[-3mm]

Butler, T. and N. Goldenfeld (2009). "Optimality Properties of a Proposed Precursor to the Genetic Code." arXiv http://arxiv.org/abs/0905.2932.\\[-3mm]

Caporaso, J. G., M. Yarus, et al. (2005). "Error Minimization and Coding Triplet/Binding Site Associations Are Independent Features of the Canonical Genetic Code." Journal of Molecular Evolution 61(5): 597-607.\\[-3mm]

Cleaves, H. J., J. H. Chalmers, et al. (2008). "A reassessment of prebiotic organic synthesis in neutral planetary atmospheres." Orig Life Evol Biosph 38(2): 105-15.\\[-3mm]

Copley, S. D., E. Smith, et al. (2005). "A mechanism for the association of amino acids with their codons and the origin of the genetic code." Proc Natl Acad Sci U S A 102(12): 4442-7.\\[-3mm]

Crick, F. H. (1968). "The origin of the genetic code." J Mol Biol 38(3): 367-79.\\[-3mm]

Darwin, C. (1859). On the Origin of Species by Means of Natural Selection, or the Preservation of Favoured Races in the Struggle for Life. London: , John Murray, Albemarle Street.\\[-3mm]

Di Giulio, M. (2005). "The origin of the genetic code: theories and their relationships, a review." Biosystems 80(2): 175-184.\\[-3mm]

Di Giulio, M. (2008). "An extension of the coevolution theory of the origin of the genetic code." Biol Direct 3: 37.\\[-3mm]

Di Giulio, M., M. R. Capobianco, et al. (1994). "On the optimization of the physicochemical distances between amino acids in the evolution of the genetic code." J Theor Biol 168(1): 43-51.\\[-3mm]

Drummond, D. A. and C. O. Wilke (2008). "Mistranslation-induced protein misfolding as a dominant constraint on coding-sequence evolution." Cell 134(2): 341-352.\\[-3mm]

Dueck, G. (1993). "New optimization heuristics: the great deluge algorithm and the record-to-record travel." Journal of Computional Physics 104: 86-92.\\[-3mm]

Freeland, S. J. and L. D. Hurst (1998). "The Genetic Code Is One in a Million." Journal of Molecular Evolution 47(3): 238-248.\\[-3mm]

Freeland, S. J., R. D. Knight, et al. (2000). "Measuring adaptation within the genetic code." Trends Biochem Sci 25(2): 44-5.\\[-3mm]

Freeland, S. J., R. D. Knight, et al. (2000). "Early Fixation of an Optimal Genetic Code." Molecular Biology and Evolution 17: 511-518.\\[-3mm]

Freeland, S. J., T. Wu, et al. (2003). "The case for an error minimizing standard genetic code." Orig Life Evol Biosph 33(4-5): 457-477.\\[-3mm]

Gilis, D., S. Massar, et al. (2001). "Optimality of the genetic code with respect to protein stability and amino-acid frequencies." Genome Biol 2(11): 49.1-49.12.\\[-3mm]

Haig, D. and L. D. Hurst (1991). "A quantitative measure of error minimization in the genetic code." Journal of Molecular Evolution 33(5): 412-417.\\[-3mm]

Higgs, P. G. (2009). "A four-column theory for the origin of the genetic code: tracing the evolutionary pathways that gave rise to an optimized code." Biol Direct 4: 16.\\[-3mm]

Higgs, P. G. and PPudritz R.E. (2009). "A thermodynamic basis for prebiotic amino acid synthesis and the nature of the first genetic code." Astrobiology. 9(5): 483-490.\\[-3mm]

Jukes, T. H. (1973). "Possibilities for the evolution of the genetic code from a preceding form." Nature 246(5427): 22-6.\\[-3mm]

Knight, R. (2001). The Origin and Evolution of the Genetic Code: Statistical and Experimental Investigations. Princeton, Princeton Univ. Ph.D.\\[-3mm]

Knight, R. D., S. J. Freeland, et al. (1999). "Selection, history and chemistry: the three faces of the genetic code." Trends Biochem Sci 24(6): 241-247.\\[-3mm]

Knight, R. D., S. J. Freeland, et al. (2001). "Rewiring the keyboard: evolvability of the genetic code." Nat. Rev. Genet 2: 49-58.\\[-3mm]

Knight, R. D. and L. F. Landweber (1998). "Rhyme or reason: RNA-arginine interactions and the genetic code." Chem. Biol 5(9): 215-220.\\[-3mm]

Knight, R. D. and L. F. Landweber (2000). "Guilt by association: The arginine case revisited." RNA 6(04): 499-510.\\[-3mm]

Knight, R. D., L. F. Landweber, et al. (2003). Tests of a Stereochemical Genetic Code. Translation Mechanism. J. Lapointe and L. Brakier-Gingras, Kluwer Academic/Plenum Publishers, New York: 115-128.\\[-3mm]

Kobayashi, K., M. Tsuchiya, et al. (1990). "Abiotic synthesis of amino acids and imidazole by proton irradiation of simulated primitive earth atmospheres." Origins of Life and Evolution of the Biosphere 20(2): 99-109.\\[-3mm]

Koonin, E. V. and A. S. Novozhilov (2009). "Origin and evolution of the genetic code: the universal enigma." IUBMB Life 61(2): 99-111.\\[-3mm]

Kramer, E. B. and P. J. Farabaugh (2007). "The frequency of translational misreading errors in E. coli is largely determined by tRNA competition." RNA 13(1): 87-96.\\[-3mm]

Massey, S. E. (2008). "A neutral origin for error minimization in the genetic code." J Mol Evol 67(5): 510-6.\\[-3mm]

Miller, S. L. (1953). "A production of amino acids under possible primitive earth conditions." Science 117(3046): 528-9.\\[-3mm]

Miller, S. L. and H. C. Urey (1959). "Organic compaund synthesis on the primitive earth." Science 130(3370): 245-251.\\[-3mm]

Miller, S. L., H. C. Urey, et al. (1976). "Origin of organic compounds on the primitive earth and in meteorites." J Mol Evol 9(1): 59-72.\\[-3mm]

Noller, H. F. (2004). "The driving force for molecular evolution of translation." RNA 10(12): 1833-1837.\\[-3mm]

Noller, H. F. (2006). Evolution of ribosomes and translation from an RNA world. The RNA World. R. F. Gesteland, T. R. Cech and J. F. Atkins. Cold Spring Harbor, NY, Cold Spring Harbor laboratory press.\\[-3mm]

Novozhilov, A. S., Y. I. Wolf, et al. (2007). "Evolution of the genetic code:  partial optimization of a random code for robustness to translation error in a rugged fitness landscape." Biology Direct 2(24).\\[-3mm]

Osawa, S. (1995). Evolution of the genetic code, Oxford University Press.\\[-3mm]

Patel, A. (2005). "The triplet genetic code had a doublet predecessor." J Theor Biol 233(4): 527-532.\\[-3mm]

Penny, D. (2005). "An interpretative review of the origin of life research." Biology and Philosophy 20(4): 633-671.\\[-3mm]

Rumer, I. B. (1966). "On codon systematization in the genetic code." Dokl Akad Nauk SSSR 167(6): 1393-1394.\\[-3mm]

Stoltzfus, A. and L. Y. Yampolsky (2007). "Amino Acid Exchangeability and the Adaptive Code Hypothesis." J Mol Evol 65(4): 456-462.\\[-3mm]

Tolstrup, N., J. Toftgard, et al. (1994). "Neural network model of the genetic code is strongly correlated to the GES scale of amino acid transfer free energies." J Mol Biol 243(5): 816-20.\\[-3mm]

Travers, A. (2006). "The evolution of the genetic code revisited." Orig Life Evol Biosph 36(5-6): 549-555.\\[-3mm]

Trifonov, E. N. (2000). "Consensus temporal order of amino acids and evolution of the triplet code." Gene 261(1): 139-151.\\[-3mm]

Trifonov, E. N. (2004). "The triplet code from first principles." J Biomol Struct Dyn 22(1): 1-11.\\[-3mm]

Vol'kenshtein, M. V. and I. B. Rumer (1967). "Systematics of codons." Biofizika 12(1): 10-13.\\[-3mm]

Woese, C. R. (1965). "On the evolution of the genetic code." Proc Natl Acad Sci USA 54(6): 1546-1552.\\[-3mm]

Woese, C. R. (1965). "Order in the Genetic Code." Proceedings of the National Academy of Sciences 54(1): 71-75.\\[-3mm]

Woese, C. R. (1967). The Genetic Code: The Molecular Basis for Genetic Expression, Harper \& Row.\\[-3mm]

Woese, C. R., D. H. Dugre, et al. (1966). "The Molecular Basis for the Genetic Code." Proceedings of the National Academy of Sciences 55(4): 966-974.\\[-3mm]

Woese, C. R., R. T. Hinegardner, et al. (1964). "Universality in the Genetic Code." Science 144: 1030-1031.\\[-3mm]

Wolf, Y. I. and E. V. Koonin (2007). "On the origin of the translation system and the genetic code in the RNA world by means of natural selection, exaptation, and subfunctionalization." Biology Direct 2(14).\\[-3mm]

Wong, J. T. F. (1975). "A Co-Evolution Theory of the Genetic Code." Proceedings of the National Academy of Sciences 72(5): 1909-1912.\\[-3mm]

Wong, J. T. F. (1981). "Coevolution of genetic code and amino acid biosynthesis." Trends Biochem. Sci 6: 33-35.\\[-3mm]

Wong, J. T. F. (2005). "Coevolution theory of the genetic code at age thirty." BioEssays 27(4): 416-425.\\[-3mm]

Wong, J. T. F. (2007). "Question 6: Coevolution Theory of the Genetic Code: A Proven Theory." Origins of Life and Evolution of Biospheres 37(4): 403-408.\\[-3mm]

Wu, H. L., S. Bagby, et al. (2005). "Evolution of the genetic triplet code via two types of doublet codons." J Mol Evol 61(1): 54-64.\\[-3mm]

Yarus, M. (1991). "An RNA-amino acid complex and the origin of the genetic code." New Biol 3(2): 183-9.\\[-3mm]

Yarus, M. (2000). "RNA-ligand chemistry: A testable source for the genetic code." RNA 6(04): 475-484.\\[-3mm]

Yarus, M., J. G. Caporaso, et al. (2005). "Origins of the Genetic Code: The Escaped Triplet Theory." Annual Review of Biochemistry 74: 179-198.\\[-3mm]

Ycas, M. (1969). The Biological Code. Amsterdam, North-Holland.

\end{document}